Phantom chain simulations for fracture of polymer networks created from star polymer mixtures of different functionalities

*Yuichi Masubuchi

Department of Materials Physics, Nagoya University,

Nagoya 4648603, JAPAN

ver Oct 25 2023

submitted to Polymer Journal

*correspondence should be addressed, e-mail: mas@mp.pse.nagoya-u.ac.jp

**Abstract**

Fujiyabu et al. have experimentally reported that mixing of 3-arm star prepolymers into 4-arm analog improves the toughness of the resultant polymer networks compared to the base network composed of 4-arm star polymers only. For the mechanism of this phenomenon, this study conducted phantom chain simulations for polymer networks composed of mixtures of star branch prepolymers with equal arm length and different arm numbers, $(f_1, f_2) = (3,4)$, $(3,6)$ and $(3,8)$, for various $f_1 = 3$ prepolymer fractions $\varphi_3$. The networks were created via end-linking reactions between prepolymers traced by a Brownian dynamics scheme, and the network structure was stored at different conversion ratios $\varphi_c$, ranging from 0.6 and 0.9. The cycle rank of the gelated networks $\xi$ is fully consistent with the mean field theory, demonstrating that the examined network structure is statistically fair. The networks were stretched with energy minimization until the break, and fracture characteristics including strain at break $\varepsilon_b$, stress at break $\sigma_b$, work for fracture $W_b$, and the ratio of broken strands $\varphi_{bb}$, were obtained. $\varepsilon_b$, $\sigma_b/\varphi_{bb}$, and $W_b/\varphi_{bb}$ data plotted against $\xi$ roughly follow the master curves reported for the base networks without mixing, implying that the change of fracture properties by the mixing of $f_1 = 3$ mainly corresponds to a decrease of $\xi$. The mixing slightly suppresses $\sigma_b/\varphi_{bb}$ and $W_b/\varphi_{bb}$ for large $f_2$ cases compared to the base networks because of a biased breakage at the network strands without extenders, which are prepolymers with only two reacted arms. The analysis for broken strands exhibited a new master curve for the $\xi$-dependence of the molecular weight of broken strands.

**Introduction**

Despite lots of attempts, the effect of node functionality on the network fracture has not been fully clarified yet. The widely investigated direction for developing tough networks is to embed multi-functional nodes[1], for which the functionality $f$, i.e., the number of connected network strands from the node, is more than 4 employing dendrimer-type linkers[2], nanoparticles[3], block copolymers[4], etc. In contrast, Fujiyabu et al.[5] have demonstrated that hydrogels made from 3-arm star polymer sols exhibit superior fracture properties to 4-arm star analogs. They also showed that the mixtures of 3 and 4-arm star prepolymers exhibit systematic improvement of the fracture properties of resultant gels according to the mixing ratio of 3-arm star polymers. Their results imply that smaller node functionality attains better mechanical properties, contradicting the strategy employing multi-functional nodes.

On the abovementioned inconsistency for the network toughness concerning $f$, a recent simulation study[6] demonstrated that the multi-functional strategy works well to realize robust networks against the conversion ratio $\varphi_c$. In contrast, the networks with fewer functionalities surpass the multi-functional analogs when $\varphi_c$ becomes high. The simulation also revealed that the dependence of the fracture characteristics, i.e., strain at break $\varepsilon_b$, stress at break $\sigma_b$, work for fracture $W_b$, and the ratio of broken strands $\varphi_{bb}$, obtained for various $f$ and $\varphi_c$, lie on master curves if $\varepsilon_b$, $\sigma_b/\varphi_{bb}$, and $W_b/\varphi_{bb}$ are plotted as functions of the cycle rank of the examined networks $\xi$.

The reported universality concerning the $\xi$-dependence of fracture characteristics could give insights into the fundamental nature of network toughness. However, since the simulation settings are significantly idealized and simplified from actual materials, necessary conditions for the universality still need to be discovered and may hardly be realized experimentally. A possible direction for this problem is additional simulations with conditions deviating from the ideal one. For instance, a recent study[7] discussed the effect of stoichiometry of prepolymers on the $\xi$-dependence of fracture characteristics. Namely, for the binary mixture of star prepolymers, the mixing ratio of two chemistries was varied to disturb the equimolar condition. The results demonstrated that the fracture characteristics, $\varepsilon_b$, $\sigma_b$, and $W_b$, decrease with increasing the magnitude of deviation from the stoichiometric condition. However, the observed deviation does not significantly disturb the universal $\xi$-dependence of fracture characteristics, implying that the change of fracture behavior according to stoichiometry is essentially due to the shift in $\xi$.

Motivated by the work of Fujiyabu et al.[5], this study examined star polymer networks made from mixtures of prepolymers with different $f$ values, $f_1$ and $f_2$ chosen at $(f_1, f_2) = (3,4)$, $(3,6)$, and $(3,8)$ for various mixing ratios of $f_1 = 3$ polymers $\varphi_3$ and conversion ratios $\varphi_c$ of end-linking reactions. The fracture characteristics systematically varied according to $\varphi_3$, being consistent with the experiment of Fujiyabu et al.[5]. Concerning the $\xi$-dependence, $\varepsilon_b$, $\sigma_b/\varphi_{bb}$, and $W_b/\varphi_{bb}$ roughly follow the master curves drawn for the base networks, implying that the change of fracture properties induced by the mixing is essentially due to the change of $\xi$. $\sigma_b/\varphi_{bb}$ and $W_b/\varphi_{bb}$ for large $f_2$ cases slightly lay below the master curves for the base networks because of the strand extenders that are prepolymers only with two reacted arms. Details are described below.

**Model and simulations**

Following the earlier studies, phantom chain simulations were performed. Since the simulation scheme and conditions are the same as the previous studies[6–8], except for mixing different $f$ prepolymers, readers familiar with the previous papers may skip this section.

Sols of star polymer mixtures with $(f_1, f_2) = (3,4)$, $(3,6)$ and $(3,8)$ were prepared with the mixing ratio of $f_1 = 3$ prepolymers $\varphi_3$ chosen at 0.25, 0.5, and 0.75 in a simulation box with periodic boundary conditions. Each arm consisted of 5 beads, and the beads were connected by non-linear springs, for which the spring constant is given as $f_{ik} = (1 - \mathbf{b}_{ik}^2/b_{\max}^2)^{-1}$. Here, $\mathbf{b}_{ik}$ is the bond vector between beads $i$ and $k$, and $b_{\max}$ is the maximum bond length chosen at 2. This non-linear spring was introduced to suppress bond breakages due to thermal fluctuations, not external deformations. The bead number density was 8, corresponding to the prepolymer concentration $c/c^* \sim 4$ concerning the overlapping concentration $c^*$. The total number of prepolymers was fixed at 1600. Note that prepared sols are mixtures of four different prepolymers: two different $f$ polymers ($f_1$ and $f_2$) and two different chemistries (A and B) for each $f$ polymers. The numbers of A and B prepolymers for each $f$ were the same.

The equilibration and gelation processes were performed with a Brownian dynamics scheme employing the 2nd-order numerical integration[9]. In the gelation process, following experiments[10–12], the reaction was allowed only at the chain ends between different chemistries. Owing to this binary nature, primary loops and higher odd-ordered loops were not generated, whereas second and higher even-ordered loops were included[13–15]. The critical distance $r_c$ and the reaction rate $p$ were chosen at 0.5 and 0.1 for the reaction. During the gelation, the simulation snapshots with the conversion ratio at $\varphi_c = 0.6, 0.7, 0.8$, and $0.9$, were stored for stretch.

For the obtained gelated networks, energy minimization was imposed by the Broyden-Fletcher-Goldfarb-Sanno method[16] with the bead displacement parameters chosen at $\Delta u = 10^{-4}$ and $\Delta r = 10^{-2}$. To the energy-minimized networks, stepwise uniaxial deformations and energy minimization were applied alternatively without Brownian motion. After each energy minimization step, the bonds longer than the critical length of breakage were removed. The critical length was chosen at $b_c = \sqrt{1.5}$, which is smaller than $b_{\max} = 2$ to realize bond breakage only due to the applied elongation. These procedures were repeated until the network percolation was eliminated toward the elongated direction.

One may argue that the stretch process can be performed with Brownian motion, as made in earlier studies[17–20]. However, if Brownian motion is turned on, the mechanical response, including fracture, strongly depends on the stretch rate due to structural relaxation and thermally-induced bond breakage [8,20]. In particular, structural relaxation occurs after every single bond breakage, and the magnitude and relaxation time tremendously grow, reflecting the large-scale network rupture in the later stage. Since the energy minimization process can skip this structural relaxation, the number of simulation parameters can be reduced, and thus, it has been employed in some earlier studies[21–23].

**Results**

Figure 1 shows typical snapshots of a network for $(f_1, f_2, \varphi_3, \varphi_c) = (3,8,0.5,0.9)$ after gelation (a), after energy minimization (b), and under stretch (c)-(f). Here, prepolymers with different $f$ are displayed in different colors. Overall, the fracture behavior is similar to what was previously reported for star-polymer networks without mixing[6–8]. The gelated structure in panel (a) is reasonably homogeneous, irrespective of mixing for the chosen segment density. In contrast, the network becomes somewhat inhomogeneous after energy minimization in panel (b) due to a structure change according to the force balance around network nodes. As the network was stretched, some strands broke, as shown in panel (g). $\varphi_u$ is the number fraction of unconnected strands, and $\varphi_u = 1 - \varphi_c = 0.1$ before stretch. $\varphi_u$ remained unchanged in $\varepsilon \lesssim 1.5$, gradually increasing with a further increase of $\varepsilon$. In this specific case, $\varphi_u$ rapidly raised at $\varepsilon \sim 2$ due to cascade bond breakage, and the network percolation was eliminated between $\varepsilon = 2.06$ and $2.07$. This bond breakage is reflected in the stress-strain relation, and the stress drops to zero, as seen in panel (h) when the network percolation is eliminated. At this final stage, the network shrunk according to energy minimization consistent with zero stress, as shown in panel (f).

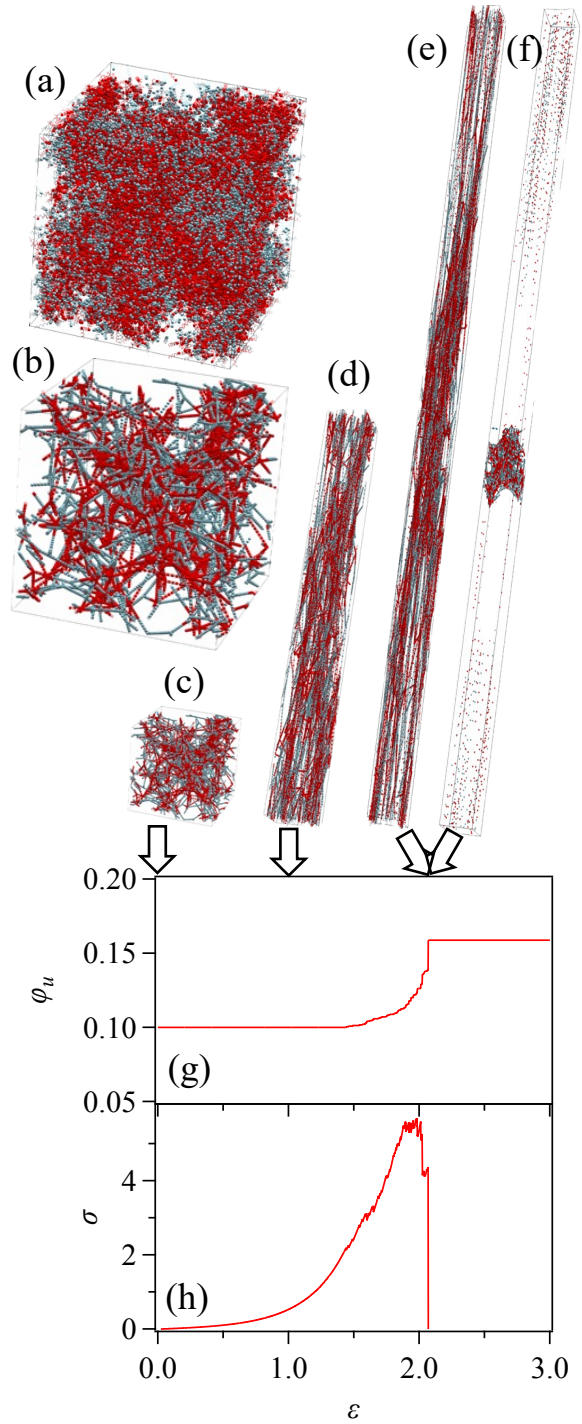

**Fig. 1** Typical snapshots for a network with $(f_1, f_2, \varphi_3, \varphi_c) = (3, 8, 0.5, 0.9)$ after gelation (a), after energy minimization (b), under stretch at uniaxial Hencky strain $\varepsilon = 0$ (c), 1.0 (d), 2.06 (e), and 2.07 (f). Segments are colored in blue and red, indicating their prepolymer functionality 3 and 8, respectively. Dispersed dots in panel (h) are isolated clusters coincidently detached from the main network. Panels (g) and (h) exhibit the development of the fraction of unconnected strands $\varphi_u$ and uniaxial stress (true stress) $\sigma$ plotted against true strain $\varepsilon$.

Figure 2 shows the stress-strain relations for $(f_1, f_2, \varphi_c) = (3,8,0.9)$ with $\varphi_3 = 0$, 0.5 and 1. Each curve corresponds to an independent simulation run. As reported earlier, $f = 3$ networks are mechanically superior to $f = 8$ analogs with $\varphi_c = 0.9$, and the mechanical properties of the mixture are located in between those for the base systems.

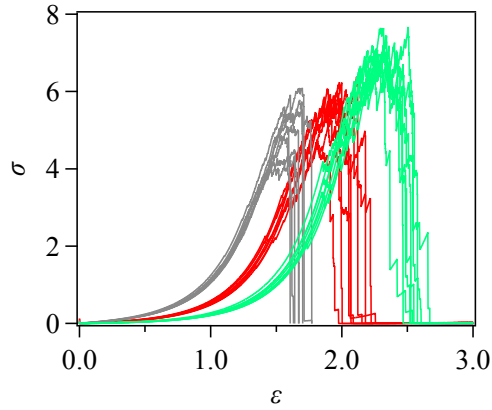

**Fig. 2** Stress-strain relations for $(f_1, f_2, \varphi_c) = (3,8,0.9)$ with $\varphi_3 = 0$ (black), 0.5 (red), and 1 (green). Each curve corresponds to an independent simulation run.

From the stress-strain relations thus observed, fracture characteristics, $\varepsilon_b$, $\sigma_b$, and $W_b$, were obtained. Note that $W_b$ was acquired by numerical integration of the stress-strain curve until the break. $\varphi_{bb}$ was also obtained as the difference between initial and final values of $\varphi_u$ (see Fig 1 (g) for example). Figure 3 shows those values plotted against $\varphi_3$ for $(f_1, f_2) = (3,4), (3,6)$ and $(3,8)$ with $\varphi_c = 0.6$-$0.9$. For all the examined cases, $\varepsilon_b$ shown in line (a) increases with increasing $\varphi_3$, and it decreases with increasing $\varphi_c$. $\sigma_b$ and $W_b$ shown in lines (b) and (c) increase with increasing $\varphi_c$, and their behavior against $\varphi_3$ depends on $\varphi_c$. Namely, with increasing $\varphi_3$, they decrease when $\varphi_c$ is small, whereas they increase with large $\varphi_c$. Concerning $(f_1, f_2) = (3,4)$ shown in the leftmost column (column 1), an increase of $\sigma_b$ and $W_b$ against $\varphi_3$ is more significant in the experimental study for tetra and tri-PEG gels[5]. This discrepancy implies that stretch-induced crystallization observed in the experiment probably plays a role in the toughness. $\varphi_{bb}$ decreases with increasing $\varphi_3$ for small $\varphi_c$, and it becomes essentially independent of $\varphi_3$ when $\varphi_c$ is large.

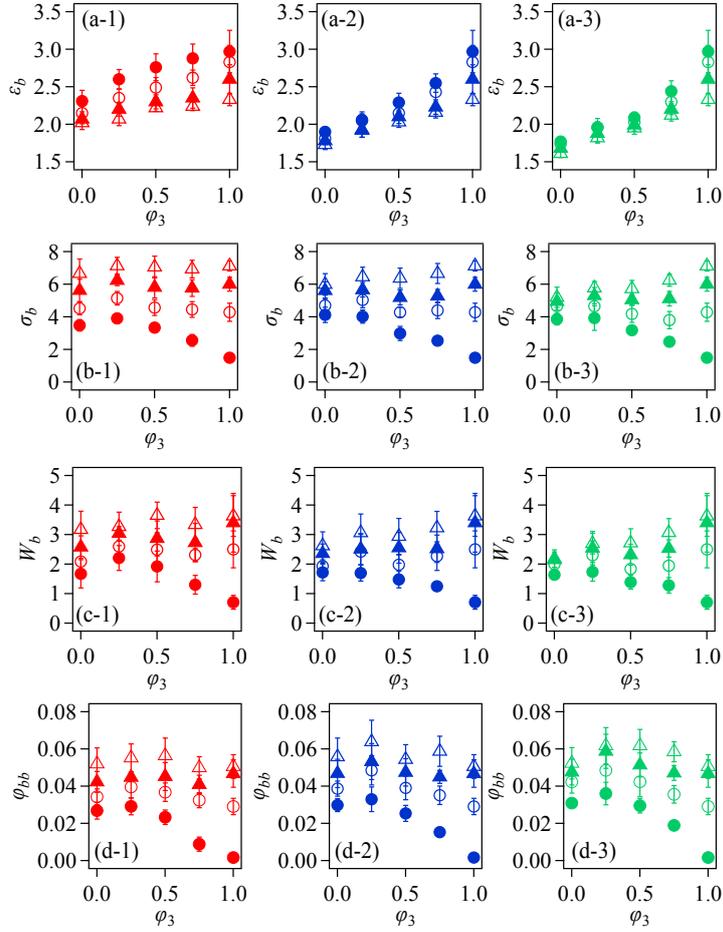

**Fig. 3** Fracture characteristics, $\varepsilon_b$ (line a), $\sigma_b$ (line b), $W_b$ (line c) and $\varphi_{bb}$ (line d) plotted against $\varphi_3$ for $(f_1, f_2) = (3,4)$ (column 1), (3,6) (column 2), and (3,8) (column 3) with $\varphi_c$=0.6 (filled circle), 0.7 (unfilled circle), 0.8 (filled triangle), and 0.9 (unfilled triangle). Error bars indicate standard deviations among eight independent simulation runs.

## Discussion

According to the previous studies[6,7], the fracture characteristics for the base networks are related to the cycle rank of the network per branch point $\xi$. Below, let us examine the results shown in Fig 3 in this aspect. Figure 4 shows $\xi$ as a function of $\varphi_3$ for $(f_1, f_2) = (3,4), (3,6)$ and $(3,8)$ with $\varphi_c$ =0.6-0.9. Here $\xi$ was obtained from the numbers of nodes and strands included in the percolated network, $\nu$ and $\mu$, as $\xi = \nu - \mu$. Note that in this definition, $\xi$ does not depend on applying the Case-Scanlan criterion[24,25]. Nevertheless, for all the examined cases, $\xi$ is fully consistent with the mean-field theory[26–28], as reported for the base networks, irrespective of the mixing. This consistency demonstrates that the end-linking reaction during gelation is essentially independent of each other in the employed simulation scheme and that the examined network structures are statistically fair.

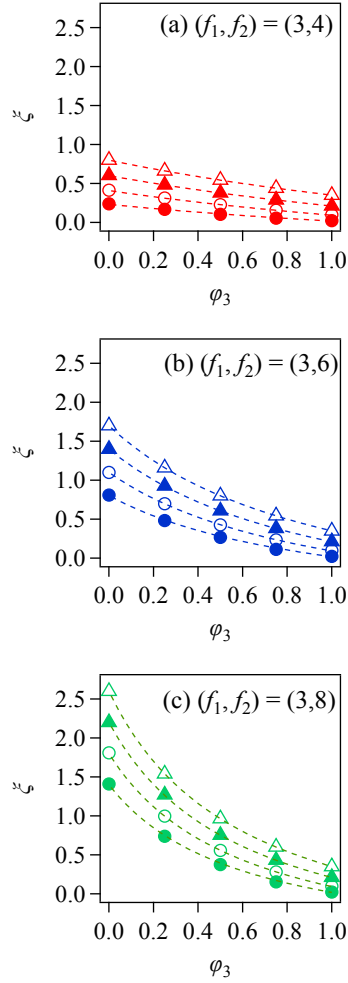

**Fig. 4** Cycle rank $\xi$ per branch point as a function of $\varphi_3$ for $(f_1, f_2) = (3,4), (3,6)$ and $(3,8)$ from top to bottom. $\varphi_c$ is 0.6 (filled circle), 0.7 (unfilled circle), 0.8 (filled triangle), and 0.9 (unfilled triangle). Broken curves are theoretical predictions according to the mean-field theory[26–28]. Standard deviations among eight independent simulation runs are within the symbols.

Figure 5 shows $\varepsilon_b$, $\sigma_b$, $W_b$, and $\varphi_{bb}$ as functions of $\xi$. $\varepsilon_b$ for the mixed networks deviate downward from the empirical master curve $\varepsilon_b = 1.85\xi^{-0.2}$ reported for the base networks[6] (red broken curve), though the deviation is insignificant compared to the error bars. Concerning the other quantities, they behave similarly against $\xi$ and $\varphi_c$. When $\varphi_c$ is small (see circles), they increase with increasing $\xi$. As $\varphi_c$ increases, the magnitude of the rise is suppressed, and they decrease with increasing $\xi$ at large $\varphi_c$ values, exhibiting the opposite trend (see triangles). These behaviors have been previously reported for the base networks[6].

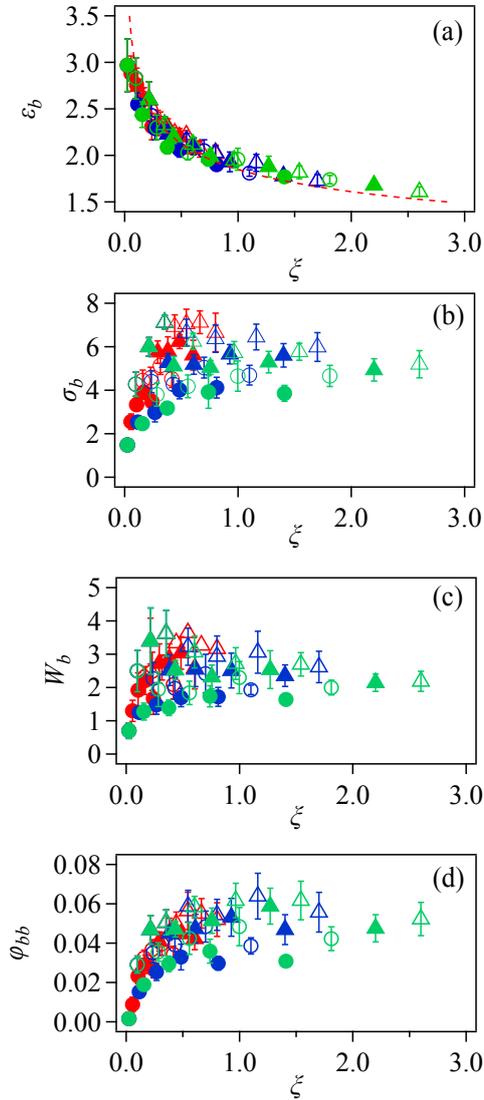

**Fig. 5** Fracture characteristics, $\varepsilon_b$ (a), $\sigma_b$ (b), $W_b$ (c) and $\varphi_{bb}$ (d) plotted against $\xi$ for $(f_1, f_2) = (3,4)$ (red), $(3,6)$ (blue) and $(3,8)$ (green) with $\varphi_c$=0.6 (filled circle), 0.7 (unfilled circle), 0.8 (filled triangle), and 0.9 (unfilled triangle). The red broken curve in panel a corresponds to the apparent relation observed in the base networks[6] written as $\varepsilon_b = 1.85\xi^{-0.2}$. Error bars indicate standard deviations among eight independent simulation runs.

Figure 6 shows $\sigma_b/\varphi_{bb}$ and $W_b/\varphi_{bb}$ plotted against $\xi$. The results for the mixed networks roughly follow the master curves reported for the base networks (shown by red broken curves), implying that the change of fracture properties induced by the mixing is mainly due to a decrease of $\xi$. Besides, one can see that the results for the mixed networks are located slightly below the master curves. This deviation means that the mixed networks are somewhat mechanically inferior to the base networks. The magnitude of the deviation becomes large as $f_2$ increases, as seen for green symbols that indicate the results for $(f_1, f_2) = (3,8)$, although insignificant compared to error bars.

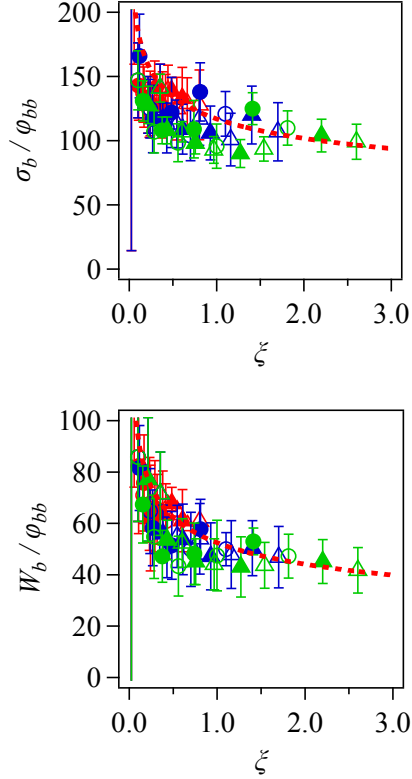

**Fig. 6** $\sigma_b/\varphi_{bb}$ and $W_b/\varphi_{bb}$ plotted against $\xi$ for $(f_1, f_2) = (3,4)$ (red), (3,6) (blue) and (3,8) (green) with $\varphi_c$=0.6 (filled circle), 0.7 (unfilled circle), 0.8 (filled triangle), and 0.9 (unfilled triangle). The red broken curves in panel a corresponds to the apparent relation observed in the base networks[6] written as $\sigma_b/\varphi_{bb} = 117\xi^{-0.2}$ and $W_b/\varphi_{bb} = 52.5\xi^{-0.25}$. Error bars indicate standard deviations among eight independent simulation runs, and some of them at small $\xi$ are large and seen as vertical lines due to small $\varphi_{bb}$ values.

The deviation of $\sigma_b/\varphi_{bb}$ and $W_b/\varphi_{bb}$ from those for the base networks is related to a bias in broken bonds. Figure 7 shows the number fraction of broken strands concerning the combination of $f$ values of the connected nodes plotted against $\varphi_3$. If the breakage occurs with a uniform probability irrespective of the combination, the fraction of broken strands should be the same as the mixing ratio of prepolymers. However, the results demonstrate that the broken strand fraction $\varphi_{b33}$, which is the number ratio of the broken strands that connect $f = 3$ nodes for both sides, is lower than the expected value, $\varphi_3{}^2$, as seen in panel (a). In contrast, the number of broken strands connected to the larger $f$ nodes, $\varphi_{b44}$, $\varphi_{b66}$, and $\varphi_{b88}$, are larger than $(1 - \varphi_3)^2$ (see panel (c)). The number of broken strands connected to different $f$ nodes (i.e., $f = 3$ in one of the ends, and $f = 4, 6$, or $8$ in the other end) is also larger than $2\varphi_3(1-\varphi_3)$ when $\varphi_3$ is large, as shown in panel (b). Note that $\varphi_c$ for the network nodes connected to the broken strands was essentially the same as for the entire system (data not shown). This biased breakage at the strands connected to large-$f$ nodes (Fig 7) explains the deviation of $\sigma_b/\varphi_{bb}$ and $W_b/\varphi_{bb}$ of mixed networks (Fig 6). In the mixed networks, $\xi$ is not uniform and is locally large around large-$f$ nodes. As reported earlier (and shown by the broken curves in Fig 6), $\sigma_b/\varphi_{bb}$ and $W_b/\varphi_{bb}$ decrease with increasing $\xi$. Thus, the biased breakage around large-$f$ nodes suppresses $\sigma_b/\varphi_{bb}$ and $W_b/\varphi_{bb}$.

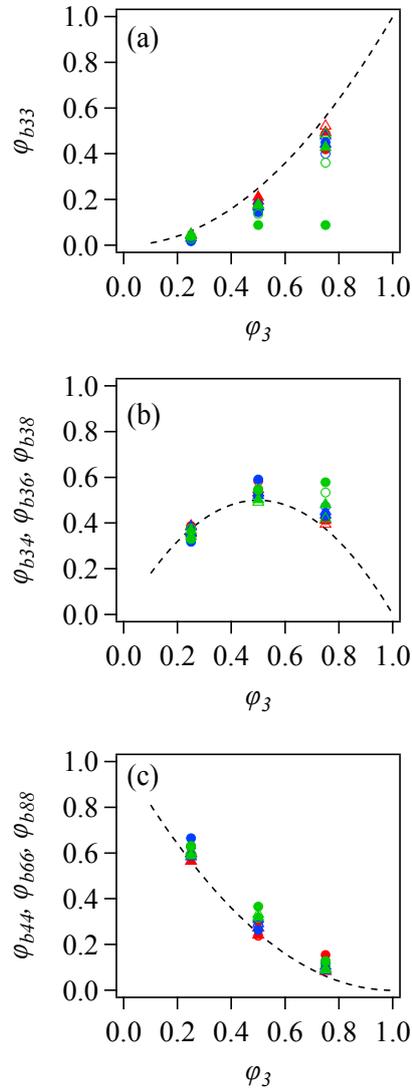

**Fig. 7** The number fraction of broken strands according to the combination of $f$ values of the network nodes at both ends. Panel (a) shows the fraction of broken strands with $f = 3$ for both sides $\varphi_{b33}$ plotted against $\varphi_3$. Panel (b) shows the fraction of broken strands connected to the nodes with different $f$ values, $\varphi_{b34}$, $\varphi_{b36}$, and $\varphi_{b38}$. Panel (c) shows the fraction of broken strands connected to large $f$ values, $\varphi_{b44}$, $\varphi_{b66}$, and $\varphi_{b88}$. Symbols correspond to the results for $(f_1, f_2) = (3,4)$ (red), (3,6) (blue) and (3,8) (green) with $\varphi_c$=0.6 (filled circle), 0.7 (unfilled circle), 0.8 (filled triangle), and 0.9 (unfilled triangle). Broken curves are expected values for the case where the probability of breakage is uniform.

A question arising here is why the biased breakage occurs around large $f$ nodes. The reason is that a large number of $f = 3$ prepolymers play a role as strand extenders[6]. Namely, prepolymers do not act as effective network nodes but extend the strand length when only two arms are reacted. Most of $f = 3$ prepolymers fall in such a condition, particularly when $\varphi_c$ is small. The breakage is suppressed for extended strands, as shown below.

Figures 8 (a)-(d) show the number-average molecular weight $M_n$ between network nodes with more than three reacted arms. Here, $M_n$ is normalized by the span molecular weight of prepolymers $M_0$, corresponding to the strand molecular weight of fully reacted networks. For the base networks without mixing with large $f$ (at $\varphi_3 = 0$), $M_n/M_0$ is close to unity even with small $\varphi_c$. In contrast, the network with $f = 3$ ($\varphi_3 = 1$) contains many strand extenders even at large $\varphi_c$, and thus, $M_n/M_0$ is larger than unity. As reported earlier[6,7], for the base networks, the change of $M_n/M_0$ according to $f$ and $\varphi_c$ can be summarized as a single master curve if $M_n/M_0$ is plotted against $\xi$ written as $M_n/M_0 = 1 + \exp(-4\xi)$, as drawn by the red broken curve shown in Fig 8 (d). Concerning $M_n/M_0$ for the mixed networks, it falls between the base systems when plotted against $\varphi_c$, as shown in panels (a)-(c). The $\xi$-dependence is similar to those for the base networks but slightly deviates from the master curve, as seen in panel (d). Namely, the data for $(f_1, f_2) = (3,4)$ (red symbols) lie on the master curve, and as $f_2$ increases, the data exhibit upward deviation from the curve. This deviation is because the value of $\varphi_c$ is smaller at the same $\xi$ for larger $f_2$, as shown in Fig 4. In the networks with smaller $\varphi_c$, there exist more strand extending $f = 3$ prepolymers. Figures (e)-(g) exhibit the number-average molecular weight taken only for broken strands $M_{bn}$. Apparently, $M_{bn}$ is smaller than $M_n$, demonstrating that the bond breakage is biased to the non-extended strands, consistent with the results shown in Fig 7. Interestingly, a master curve can be drawn for $M_{bn}/M_0$ as a function of $\xi$ as $M_n/M_0 = 1 + 0.5\exp(-4\xi)$, though no theoretical explanation can be given.

Concludingly, the downward deviation of $\sigma_b/\varphi_{bb}$ and $W_b/\varphi_{bb}$ for the mixed networks seen in Fig 6 can be explained as follows. The results in Fig 8 demonstrate that bond breakage occurs with a bias to the short strands without extending prepolymers. Because prepolymers with large $f$ rarely become strand extenders, the breakage occurs at these polymers with a bias, as shown in Fig 7. Consequently, the fracture characteristics are suppressed as those for large-$f$ base networks, whereas $\xi$ increases with increasing $\varphi_3$.

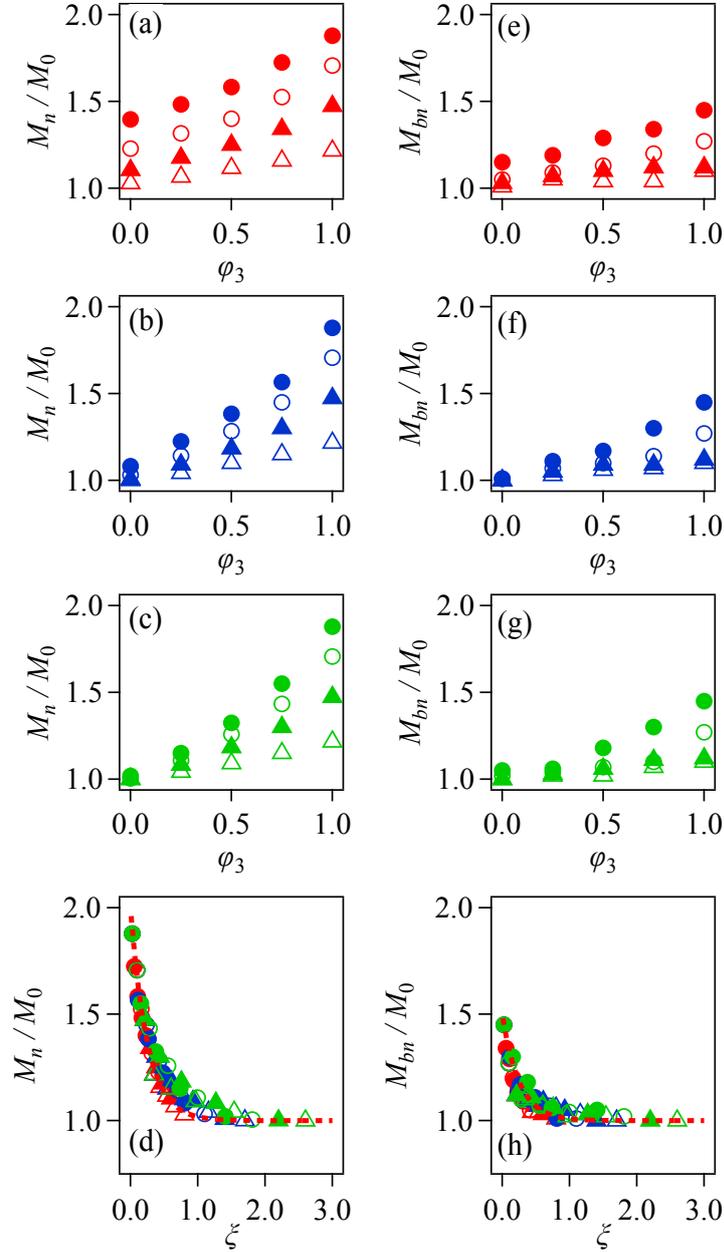

**Fig. 8** Number-averaged molecular weight $M_n$ between network nodes with more than 3 reacted arms (a-d) and that for broken strands $M_{bn}$ (e-h) normalized by that for the fully reacted network $M_0$. In panels (a)-(c) and (e)-(g), the data for $(f_1, f_2) = (3,4)$, $(3,6)$, and $(3,8)$ are separately shown from top to bottom as a function of $\varphi_3$, whereas all the data are summarized as a function of $\xi$ in panels (d) and (h). The data for $(f_1, f_2) = (3,4)$, $(3,6)$, and $(3,8)$ are shown in red, blue, and green, respectively. Filled circle, unfilled circle, filled triangle, and unfilled triangle indicate the data with $\varphi_c$ at 0.6, 0.7, 0.8, and 0.9, respectively. Red broken curves in panels (d)

and (h) show apparent relationships written as $M_n/M_0 = 1 + \exp(-4\xi)$ and $M_{bn}/M_0 = 1 + 0.5\exp(-4\xi)$.

**Conclusions**

Phantom chain simulations were performed for polymer networks made from mixtures of star polymers with different arm numbers diverging from branch points, $f_1$ and $f_2$, with $(f_1, f_2) = (3,4)$, $(3,6)$, and $(3,8)$. The arm molecular weight is monodisperse. The networks were created via gelation according to the end-linking reactions between star prepolymers simulated by the Brownian dynamics scheme with various fractions of $f_1 = 3$ prepolymer $\varphi_3$ until the conversion ratio $\varphi_c$ reached at certain values. The cycle rank $\xi$ of these gelated networks is fully consistent with the mean-field theory, demonstrating that the examined networks are statistically fair. The networks were stretched with energy minimization and fracture characteristics, i.e., strain at break $\varepsilon_b$, stress at break $\sigma_b$, work for fracture $W_b$, and the ratio of broken strands $\varphi_{bb}$, were obtained. $\varepsilon_b$ monotonically increases with increasing $\varphi_3$ irrespective of $\varphi_c$. $\sigma_b$ and $W_b$ also increase with increasing $\varphi_3$ when $\varphi_c$ is large. This change of fracture characteristics by mixing $f_1 = 3$ prepolymers is consistent with experiments for $(f_1, f_2) = (3,4)$ by Fujiyabu et al.[5] In contrast, $\sigma_b$ and $W_b$ decrease when $\varphi_c$ is small, though not explored earlier. These fracture characteristics concerning $(f_1, f_2)$, $\varphi_3$, and $\varphi_c$, were summarized as functions of $\xi$. The $\xi$-dependence of $\varepsilon_b$, $\sigma_b/\varphi_{bb}$ and $W_b/\varphi_{bb}$ roughly follow the master curves for the base networks without mixing, implying that the change of fracture properties induced by the mixing is mainly due to the change of $\xi$. $\sigma_b/\varphi_{bb}$ and $W_b/\varphi_{bb}$ slightly lie below the master curves for the base networks when $f_2$ is large, demonstrating that the network toughness is somewhat reduced by the mixing. The reason for this phenomenon was analyzed in terms of the strand extenders that are prepolymers only with two reacted arms. The results revealed that the breakage occurs with a bias to the strands without such extenders. Reflecting this bias, $\sigma_b/\varphi_{bb}$ and $W_b/\varphi_{bb}$ of mixtures are close to those for the base networks with large-$f$, whereas $\xi$ is suppressed by the mixing of $f_1 = 3$ prepolymers. Besides, a new master curve was found for the molecular weight of broken strands $M_{bn}$ plotted against $\xi$.

The results, including those reported in the previous studies, demonstrate the capability of the analysis according to $\xi$ that can summarize fracture characteristics obtained for various conditions. However, the explored conditions have still been limited, and mapping to experimental systems is not straightforward. Supplemental studies for various systems, including star polymer networks with different prepolymer concentrations and molecular weights and networks made from mixtures of linear and star prepolymers, are ongoing, and the results will be published elsewhere.